\documentclass[fleqn,twoside]{article}
\usepackage{espcrc2}
\usepackage{amssymb,amsfonts,amsmath}

\usepackage{graphicx}
\usepackage{epsfig}
\usepackage[figuresright]{rotating}

\newcommand{\eq}{\begin{equation}}
\newcommand{\qe}{\end{equation}}
\newcommand{\ear}{\begin{eqnarray}}
\newcommand{\rae}{\end{eqnarray}}
\newcommand{\Z}{\mathbb{Z}}
\newcommand{\bra}{\langle}
\newcommand{\ket}{\rangle}
\newcommand{\um}{\frac12}
\newcommand{\uq}{\frac14}

\title{ The phase diagram of the three-dimensional  $Z_2$ gauge Higgs 
system\\ at zero
  and finite temperature\thanks{Based on a poster by A.Rago and a talk by F.Gliozzi}
       }

\author{Luigi Genovese, Ferdinando Gliozzi, Antonio Rago and Christian
  Torrero \address[torino]
{Dipartimento di Fisica Teorica, Universit\`a di Torino and INFN,
sezione di Torino, via P. Giuria, 1, I-10125 Torino, Italy.}}

\begin{document}

\begin{abstract}
We study the effect of adding a matter field  to the $\Z_2$ gauge model in 
three dimensions at zero and  finite temperature.  
Up to a given value of the parameter regulating the coupling, the
matter field produces a slight shift of the transition line without 
changing the universality class of the pure gauge
theory, as seen by finite size scaling analysis or by comparison, in
the finite temperature case, to exact formulas of conformal field
theory.
At zero temperature the critical line turns into a first-order
transition.
The fate of this kind of transition in the finite temperature case is 
discussed. 

\end{abstract}

\maketitle

\section{INTRODUCTION}
The three-dimensional $\Z_2$ gauge Higgs system is perhaps the simplest 
example of a gauge theory coupled to a matter field. 
Its action can be written as
\begin{equation}
S=-\beta_G\sum_{\square\in\Lambda}U_{\square}
-\beta_I\sum_{\bra xy\ket}\sigma_xU_{xy}\sigma_y
\end{equation}
where $U_{\square}=\prod_{\ell\,\in\square}U_\ell$,
$U_\ell=\pm1$, $\Lambda$ is a cubic lattice and $\sigma=\pm1$ denotes 
the matter field. This model is self-dual under a Kramers-Wannier 
transformation: 
\begin{equation}
Z(\beta_G,\beta_I)=\sum_{conf.}{\rm e}^{-S}\propto Z(\tilde{\beta}_I,
\tilde{\beta}_G)~,
\label{duality}
\end{equation}
with $\tilde\beta=-\log\sqrt{\tanh\beta}$. Its phase
structure has been determined long ago \cite{Jongeward:1980wx} and it has been shown
to be very similar to that of $SU(2)$ gauge system coupled to a matter
field in the fundamental representation, but of course it is much
simpler, moreover the coupling to the Ising matter can be now
efficiently implemented by a non-local cluster algorithm \cite{sw}  
which leads to very accurate Monte Carlo (MC) simulations. Therefore it appears
as an ideal laboratory to test new ideas on the confining-deconfining
properties of coupled gauge models. Recently it has been used to study
the phenomenon of string breaking 
\cite{Gliozzi:2000yg,Gliozzi:2001tu,Gliozzi:2002pd} in order to probe
a general mechanism proposed to explain why this phenomenon is not
visible in the Wilson loops \cite{Gliozzi:1999wq}.

In this contribution we report an accurate analysis
of the transition lines of this model at zero and at finite
temperature. 

At zero temperature we use the histogram method and the 
re-weighting techniques \cite{fm,ma,fs} to locate the first order 
transition line and the standard finite size scaling  study of 
the continuous transitions, which turn out to be in the universality 
class of $3D$ Ising model. We find that the apparent triple point
suggested by the old analysis based on the hysteresis 
cycles \cite{Jongeward:1980wx} is a finite size effect.

At finite temperature we argue that the matter field term in the
action, for not too large values of $\beta_I$, is an irrelevant 
operator of the deconfining transition of the pure
$\Z_2$ gauge theory, which is known to belong to the $2D$ Ising
universality class. We support this conjecture by comparing the
Polyakov correlator with the exact formula of spin-spin correlator 
in a finite box given by the underlying 2D conformal theory. This
allows us to locate  the transition 
lines in the plane $\beta_G,\beta_I$ . For large enough $\beta_I$ the 
nature of the transitions changes
and becomes strongly influenced by the boundary conditions. 

\section{ZERO TEMPERATURE}
The problem of detecting a first order transition by MC simulations on a finite
system of volume $V=L^3$ can be solved by computing the histogram of 
energy distribution $P(E,L)$ at a point $\beta_G,\beta_I$ close to the
transition line and then extrapolating the data to nearby values 
\cite{fm,ma,fs}
\begin{figure}[t]
\includegraphics[scale=.6]{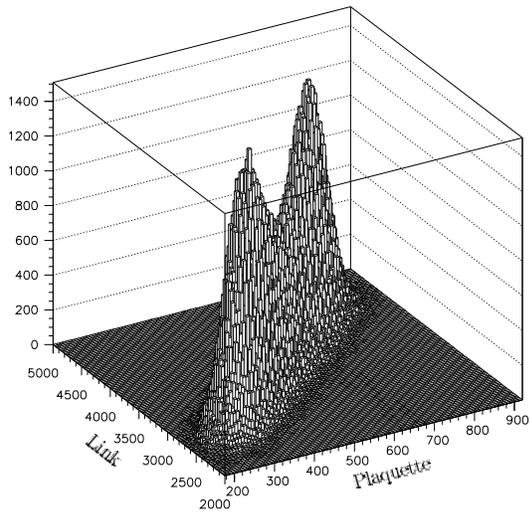}
\caption{The double peak structure of the histogram of the link and
  the plaquette distribution near a first-order transition.}
\label{fig:peaks}
\vspace{-1mm}
\end{figure}
\begin{figure}[htb]
\includegraphics[scale=.3]{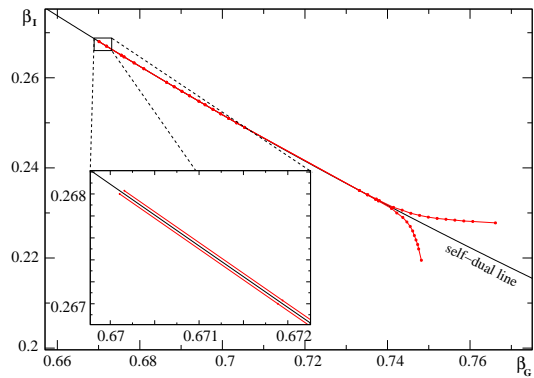}
\caption{ Location of the double peak in the L=18 lattice as a result
  of nine actual simulations corresponding to $1.2\,10^7$ MC steps.}
\label{fig:pezzo4}
\vspace{-1mm}
\end{figure}
\begin{figure}[htb]
\includegraphics[scale=.6]{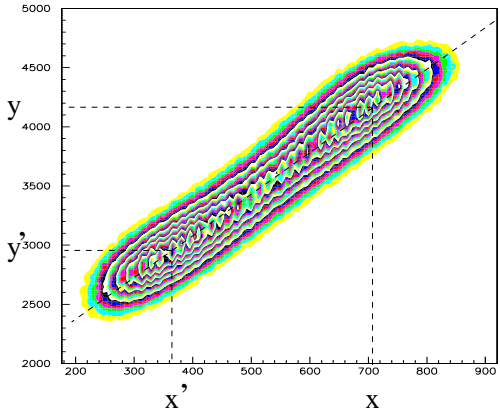}
\caption{Contour plot of the double peak structure of
  Fig.(\ref{fig:peaks}) in the plaquette ($x$) and link ($y$) plane.}
\label{fig:levels}
\vspace{-1mm}
\end{figure}
\begin{figure}[htb]
\includegraphics[scale=.3]{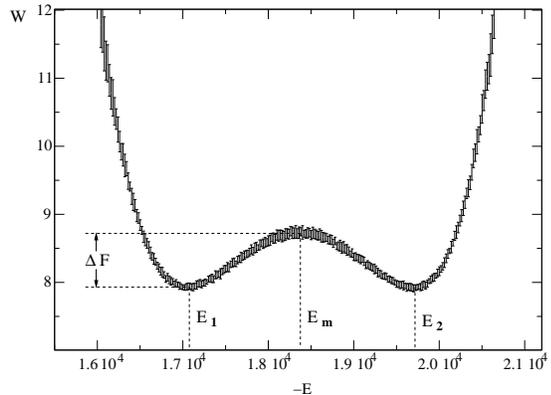}
\caption{A typical plot of $W=-\log P(E,L)$ with $L=18$ on the
  self-dual line.}
\label{fig:free_energy}
\vspace{-1mm}
\end{figure}
\begin{figure}[htb]
\includegraphics[scale=.3]{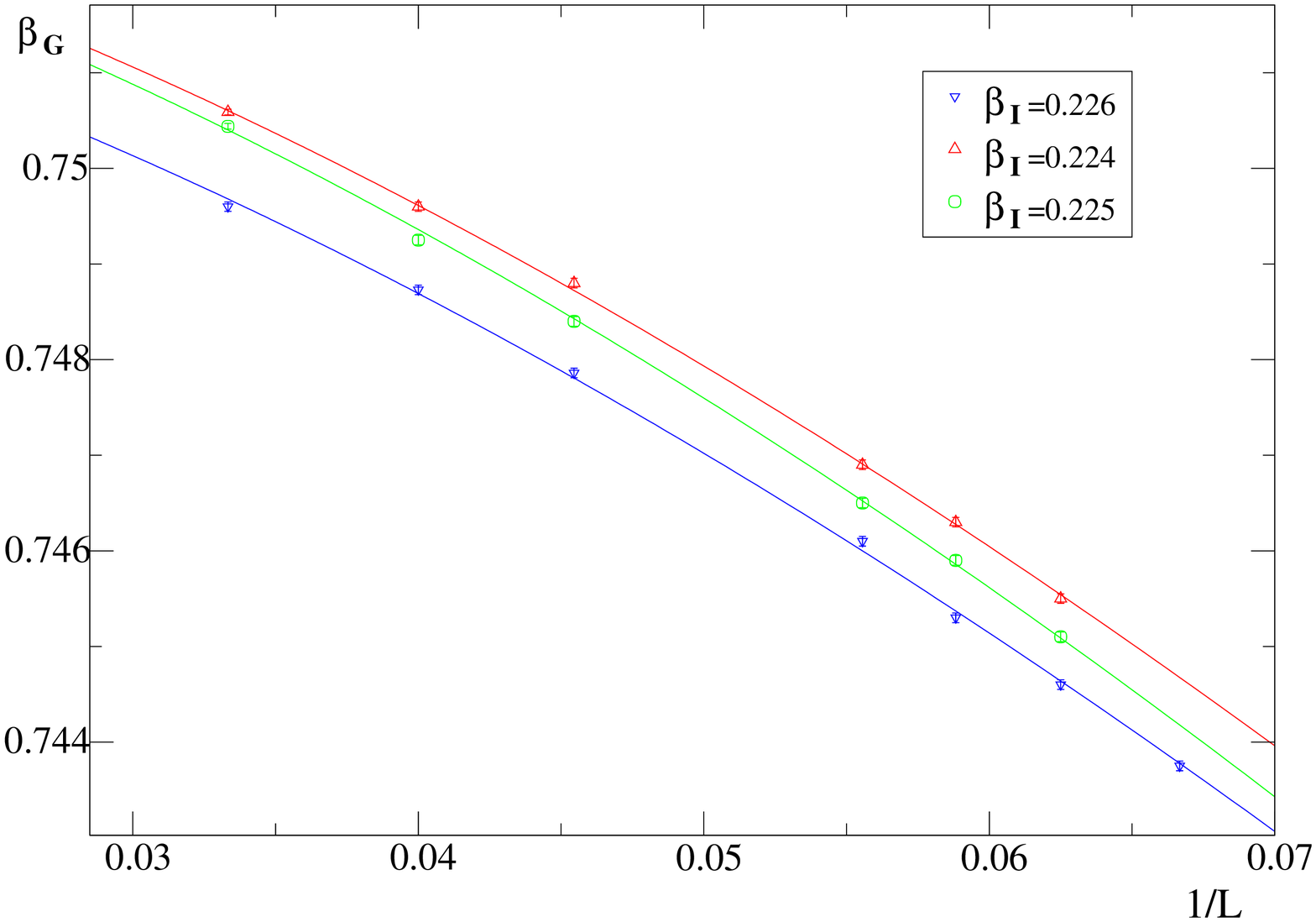}
\caption{Finite size scaling behavior of the transition line near the 
bifurcation. The three curves are the fits to Eq.(\ref{FSS}). At each
value of $L$ the points are obtained by re-weighting the data of a
single simulation.}
\label{fig:scaling}
\vspace{-1mm}
\end{figure}
\begin{figure}[htb]
\includegraphics[scale=.62]{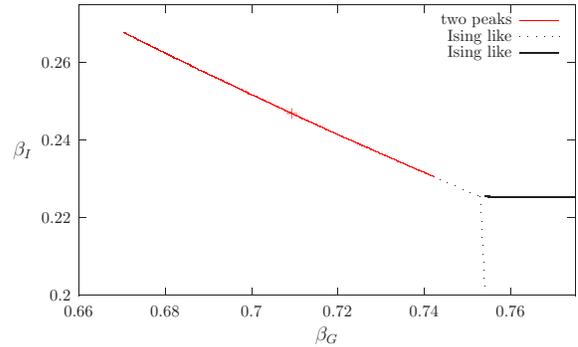}
\caption{A schematic view of the phase diagram of the $\Z_2$ gauge
  Higgs model at zero temperature  in the infinite volume limit. The
  cross indicates the point where the first-order transition has its
  maximal strength.}
\label{fig:phaslim}
\vspace{-1mm}
\end{figure}
\begin{equation}
P(E,L)=\Omega(E,L)\frac{\exp{-\beta_G E}}{Z}~~,
\label{hystogram}
\end{equation} 
where $Z$ is the partition function and $\Omega(E,L)$ is the number of
states of energy 
$E=-\sum_\square U_\square-\frac{\beta_I}{\beta_G}
\sum_{\bra xy\ket}\sigma_xU_{xy}\sigma_y$. In the vicinity of a
first order transition it
has a characteristic double peak structure as shown in the double
histogram of the links and plaquettes of  Fig.(\ref{fig:peaks}). 
A suitable re-weighting through Eq.(\ref{hystogram}) yields the line 
$\beta_I=f(\beta_G,L)$ where the two peaks 
at $E_1(\beta_G,\beta_I,L)$ and $E_2(\beta_G,\beta_I,L)$ are of equal 
height. We located numerically this line for cubic lattices of sides 
ranging from $L=10$ to $L=30$. The autocorrelation times for larger
lattices were too large in our canonical MC simulations. Perhaps more
refined techniques such as multi-canonical algorithm \cite{bn} could
reduce this correlation significantly. A typical plot of 
$\beta_G=f(\beta_I,L)$ is reported in Fig.(\ref{fig:pezzo4}): it is
formed by two dual lines which cross each other on a self-dual point. Above the
crossing point the distance of these two curves from the self-dual
line (SDL) decreases rapidly when $L$ increases and is already microscopic
at $L=18$ (see the inset of Fig.(\ref{fig:pezzo4})). If the system 
had a true triple point, one should see a triple peak near the crossing 
point, while we  observe in all the cases a sharp double peak
structure as represented
in Fig.(\ref{fig:peaks}). Denoting by $(x,y)$ and $(x',y')$ the
coordinates of the two peaks in the $(plaquette,link)$ plane 
(see Fig.(\ref{fig:levels})) it is
easy to prove that, if two states coexist at a self-dual point, they 
are related, in the thermodynamic limit, by
\ear
x+y'\,\sinh 2\beta_I=\cosh 2\beta_I\\
x'+y\,\sinh 2\beta_I=\cosh 2\beta_I
\rae 
which turn out to be approximately verified also in the finite
lattices. It is worth observing that the presence of a double peak is
not sufficient to assure a true first order transition. A useful
quantity in this regard is the bulk free-energy barrier
$\Delta F$ between the two coexisting states, defined by
  \begin{equation}
\Delta F(L)=W(E_m)-W(E_1)~,
\end{equation}
where $W=-\log P(E,L)$ and $E_m((\beta_G,\beta_I,L)$ is the local maximum
which separates the two dips at $E_1$ and $E_2$ when $W(E_1)=W(E_2)$,  
as shown in Fig.(\ref{fig:free_energy}). At a continuous transition, 
$\Delta F(L)$ is independent of $L$ and at a first-order transition it
increases monotonically with $L$. For large enough $L$ one has
\cite{b}, for periodic boundary conditions, $\Delta F(L)\sim 2\sigma
L^2$, where $\sigma$ is the interface 
tension. In our data at fixed $L$ $\Delta F$ is maximal at the
crossing point. Extrapolating to large $L$ we locate the point where
the first-order transition has its maximal strength at
$\beta_G=0.708(2)$, $\beta_I=-\log(\tanh(\beta_G))/2$ and the
corresponding interface tension is $2\sigma=.00108(9)$.

Using the behavior of $\Delta F(L)$ as a criterion for discriminating
the order of the  transition, we can prove the first-order nature only for a
small interval around the crossing point. Near the bifurcation, where 
the transition lines go off the SDL, even if  small
lattices show still the double peak structure, the transition is
second order. To extract an estimate for the infinite-volume
transition line in this region, we tried the standard finite-size
scaling form
\begin{equation}
\beta_G(L)=f(\beta_I,\infty)-c(\beta_I)\,L^{-\frac1\nu}~,
\label{FSS}
\end{equation}
which turns out to fit well the data, as Fig.(\ref{fig:scaling})
shows, using $\nu=0.63$, which is the value of
the corresponding critical index of
the  $3D$ Ising universality class. The resulting phase diagram in the
thermodynamic limit is reported in Fig.(\ref{fig:phaslim}).

\section{FINITE TEMPERATURE}

  The universality class of a continuous deconfining transition of a
pure gauge theory in $D$ dimensions is well understood in terms of the 
Svetitsky and Yaffe (SY) conjecture \cite{Svetitsky:1982gs}: 
it coincides with that of the spin
model in $D-1$ dimensions with a global symmetry coinciding with the
center {$C(G)$ of the gauge group.
What is the effect of adding matter to a pure gauge system?
there is no general answer.  Even when the  matter can be treated as a
'small' perturbation  ( using for instance the inverse mass $1/m$ as a 
perturbing parameter), it does not change the nature of the transition only 
if it is an irrelevant operator.

At first sight one is tempted to conclude that the matter acts always
as a relevant operator, since it breaks explicitly the global center
symmetry of the pure gauge theory at finite temperature, which is the 
hart of the SY conjecture.  
 
In the model at hand, when the coupling $\beta_I$ of the matter is
 small enough, it is easy to show that this is not actually the case:
the addition of the matter does not modify the universality class of
 the deconfining transition both at zero and at finite temperature. 
 The argument goes as follows. Performing a duality
 transformation as defined in Eq.(\ref{duality}), when 
$\tilde{\beta}_I\to\infty$ we recover the usual 3D Ising model. For
large $\tilde{\beta}_I$ it is possible to do a perturbation expansion in 
${\rm e}^{-\tilde{\beta}_I}$. The first order correction is due to a
 single anti-ferromagnetic link, and the corresponding  change in the
 free energy is proportional to ${\rm e}^{-4\tilde\beta_I}
\bra \sum_{\bra xy\ket}{\rm e}^{-2\tilde\beta_G\sigma_x\sigma_y}\ket$. 
Near the critical point, this may be expanded in a sum of scaling  operators
 all of which will be even under spin reversal. The dominant term is
 therefore proportional to the energy operator of the unperturbed Ising
 model, both at zero and at finite $T$. Therefore the only effect is
 that of a slight shift of the transition line, without changing
 the universal critical properties. So, in a sense, the matter field acts as
 an irrelevant perturbation of the universality class of the pure gauge
 theory. In order to extend this property to larger values of $\beta_I$
 we have to resort to numerical work.
 
At finite $T$, where the deconfining transition is known to be well
described by the $2D$ Ising universality class, we can support this
property even at larger values of $\beta_I$
 by accurate numerical tests of comparison with the exact finite size 
formulas dictated by the $2D$ conformal field theory.

At $\beta_I=0$, $N_t=6=1/T_c$  the deconfining transition is estimated
to be at $\beta_G=0.746035$ \cite{Caselle:1995wn}.
Here the Polyakov loop correlators, according to the SY
conjecture combined with the universal finite size effects 
dictated by the 2d conformal field theory, should be given by 
\cite{DiFrancesco:ez}
\begin{equation} 
\bra L(0)L(\vec{x})\ket\propto
\frac{\sum_{\nu=1}^4\vert\theta_\nu(\um z,\tau)\vert
\left\vert\frac{\theta_1'(0,\tau)}{\theta_1(z,\tau)}\right\vert}
{\sum_{\nu=2}^4\vert\theta_\nu(0,\tau)\vert},
\label{confcorr}
\end{equation}
where $\hat t$ is the temporal direction, 
$L(\vec x)=\prod_{n=1,\dots N_t}U_{\vec
  x+(n-1)\hat{t},\vec{x}+n\hat{t}}~$ is the Polyakov loop along 
$\hat t$ and $\theta_\nu(z,\tau)$  denotes  the Jacobi theta functions of
 argument $z$ and modulus $\tau=i\frac{N_y}{N_x}\;$; a point 
$\vec{x}= (x_1,x_2)$ in the spatial plane is mapped to a complex
number through $z=x_1+\tau x_2$. Note that
$
\bra L(0)L(\vec{x})\ket \sim_{\vec{x}\to0}\frac1{\vert\vec{x}\vert^\uq}~,
$
as expected at the critical point in the infinite volume limit.

If the matter field behaves as an irrelevant perturbation, this 
property should be valid even at $\beta_I>0$ at an appropriate value of 
$\beta_G(\beta_I)$.  This has been checked accurately on large lattices at
$N_t=6$ and $20\leq N_x,N_y\leq 160$.

A typical fit is reported in 
Fig.(\ref{fig:polya}). Note that Eq.(\ref{confcorr}), being a formula
 derived in the context of the conformal field theory, is valid only
 in the continuum limit, so at short distance it is expected that
 Polyakov loop correlators may be affected by lattice artifacts. Our
 data suggest that the value predicted by  the continuum limit is
 reached already at distances of $\sim$ 3 lattice spacings.
 Finite-size effects 
at criticality are rather strong due to scale invariance, and
 nontrivial. Therefore they are ideally suited to compare theoretical
 predictions with MC simulations. In particular Eq.(\ref{confcorr})
 produces strong, universal shape effects through the modulus
$\tau$, which takes into account the asymmetry of the lattice. A fit 
of the Polyakov correlators in an asymmetric lattice is reported in 
Fig.(\ref{fig:polyas}). Similar shape effects in the pure $\Z_2$ gauge theory
 at criticality where already observed in Ref. \cite{Gliozzi:1997yc}.  
\vskip -.5 cm
\begin{figure}[htb]
\includegraphics[scale=.55]{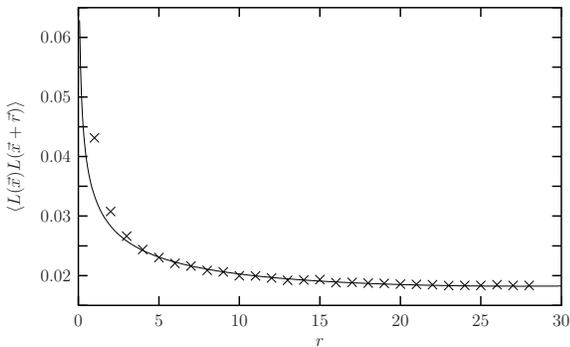}
\caption{The Polyakov loop correlator at
  $\beta_G=0.7445,~\beta_I=0.17$ in a lattice of size $6\times
  55\times 55$ compared with the critical
2D spin spin Ising correlator in a square box.}
\label{fig:polya}
\vspace{-1mm}
\end{figure}

\begin{figure}[htb]
\includegraphics[scale=.5]{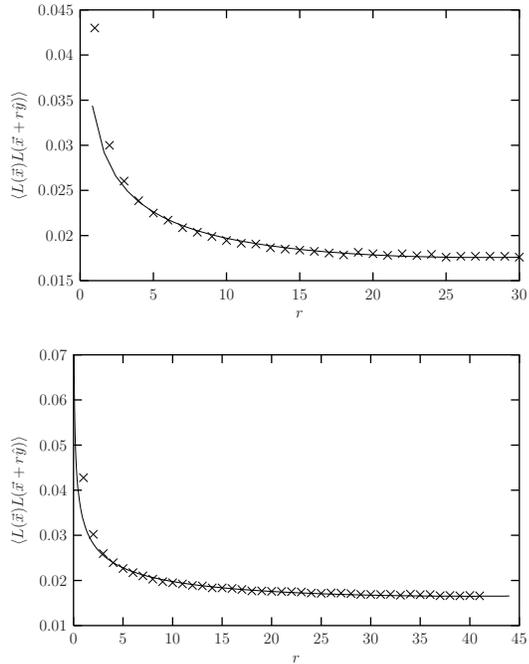}
\caption{Shape effects on the Polyakov loop correlator: same as 
Fig.(\ref{fig:polya}) with an asymmetric lattice of size $6\times
  55\times 80$. In the two figures the correlator is taken along the
  two different coordinate axes. }
\label{fig:polyas}
\vspace{-1mm}
\end{figure}
\vskip -.5 cm
We used the goodness of the data fits  to Eq.(\ref{confcorr}) 
 as a criterion  to locate the  transition line $\beta_G=f(\beta_I)$ in
the plane $\beta_G,\beta_I$ below the SDL.
The Kramers- Wannier transformation
generates another critical line which is the dual $\tilde\beta_I=
\tilde f(\tilde\beta_G)$ of the previous line (see Fig.(\ref{fig:phasediag})).
In the limit $\beta_I\to 0$ Eq.(\ref{duality}) yields
$
Z(\beta_G,0)\propto\sum_{x,y,z}
Z_{xyz}^I(\tilde\beta_G)
$
where $x,y,z$ denote periodic or anti-periodic boundary conditions (BC) in the
 3 directions, and  $Z_{xyz}^I$ is the usual Ising partition
 function. Anti-periodic BC are implemented by closed surfaces of
 anti-ferromagnetic links wrapped around the periodic directions.

When $\beta_I>0$ the sign of the links and consequently the BC 
become dynamical degrees of freedom, however
\begin{figure}[htb]
\includegraphics[scale=.55]{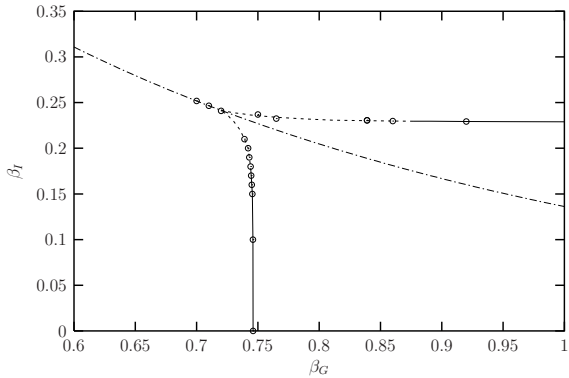}
\caption{The phase diagram at finite temperature. The points  below the
  SDL are obtained through comparison with Eq.(\ref{confcorr}). The
  points above are obtained by comparison with the critical partition
  function of  $2D$ Ising model, as explained in Ref.\cite{Caselle:1995wn}.}
\label{fig:phasediag}
\vspace{-1mm}
\end{figure}
\begin{figure}[htb]
\includegraphics[scale=.5]{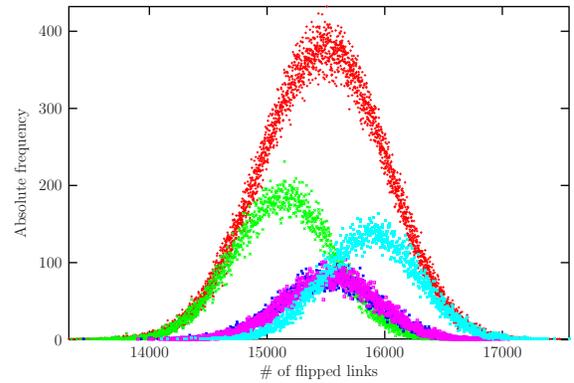}
\caption{The histogram of the link distribution (highest curve) is
separated into the contributions associated to different BC on the
transition line near $\beta_G=0.86~$.} 
\label{fig:lnkisto}
\vspace{-1mm}
\end{figure}
local updating algorithms on the critical line do not mix periodic
and  anti-periodic BC. Therefore the system behaves near the transition
line above the SDL as a pure $2D$ critical Ising model.
In the Fortuin Kasteleyn (FK) random cluster description we can
indirectly evaluate
the status of the BC by looking for the FK clusters with a linkage along the 
periodic directions.
Transitions between periodic and anti-periodic BC are possible for
not too large values of $\beta_G$.  
It turns out that when these kinds of transitions become statistically
relevant, the nature of the transition line seems modified. In
particular, the agreement of the dual transition below the SDL
 worsens and the expectation value of the link on the transition 
line above the SDL is somewhat influenced by the BC: even if the
histogram of the distribution of the link 
(or the plaquette) variable does not show any macroscopic double peak 
structure, we can  separate this distribution in various sets,
according to the linking properties of the largest FK cluster, which
gives an indirect information on the 
BC of the underlying Ising model. The result of this separation is
reported in Fig.(\ref{fig:lnkisto}), which seems to indicate a very
week first-order transition driven by the BC.    




 \end{document}